\documentclass[letterpaper,10pt,conference]{ieeeconf}

\IEEEoverridecommandlockouts
\makeatletter
\def\input@path{{./}}
\makeatother

\usepackage[T1]{fontenc}
\usepackage[utf8]{inputenc}
\usepackage{graphicx}
\graphicspath{{./}}
\usepackage{booktabs}
\usepackage{multirow}
\usepackage{array}
\usepackage{amsmath}
\usepackage{amssymb}
\usepackage{cite}
\usepackage{tikz}
\usepackage{pgfplots}
\usepackage{pgfplotstable}
\usepackage{url}

\pgfplotsset{compat=1.18}
\usetikzlibrary{arrows.meta,positioning,fit,backgrounds,shapes.geometric}

\title{\LARGE \bf
LLM-Driven CI/CD Workflow Intelligence for Cyber Systems Engineering
}

\author{Bonan Shen$^{1}$, Jiazhou Gao$^{2}$ Tao Ning$^{3}$, Wei-Jung Huang$^{4}$ and Xin Liu$^{5}$ 
\thanks{$^{1}$Independent Researcher  {\tt\small shenbonan2@gmail.com}}
\thanks{$^{2}$Independent Researcher  {\tt\small gjz140103@gmail.com}}
\thanks{$^{3}$Independent Researcher  {\tt\small ntgd1102@gmail.com}}
\thanks{$^{4}$Independent Researcher  {\tt\small william.wj.huang@gmail.com}}
\thanks{$^{5}$Independent Researcher  {\tt\small iamxinliu@gmail.com}}
}

\begin{document}

\maketitle
\thispagestyle{empty}
\pagestyle{empty}

\begin{abstract}
CI/CD workflows have become executable operational policy: they decide what gets built, tested, released, and deployed, and they mediate how maintainers interact with delivery infrastructure. That makes them an important measurement point for cyber-systems engineering. Recent large language model (LLM) work shows that workflow stages can be recognized directly from configuration files, but stage labels alone do not tell us whether a workflow is brittle, unusual for its ecosystem, or worth revising first. We present an LLM-based CI/CD analysis pipeline that combines repository enrichment, anti-pattern detection, stage mining, and recommendation generation over a large GitHub corpus. Starting from 59,550 repositories with at least 1,000 stars, we identify 34,225 projects with CI/CD and collect 127,559 configuration files. Across 75,201 analyzed workflows, the anti-pattern detector reports 434,769 findings, dominated by reliability and maintainability issues. Across 59,906 configurations, stage usage differs significantly by language ($\chi^2 = 4168.88$, $p < 0.001$, Cramer's $V = 0.063$), and domain analysis shows distinct operational profiles, including higher release and cache usage in mobile projects. For repository-level recommendation generation, few-shot prompting performs best overall, averaging 8.25 recommendations per repository with 96.1\% YAML-valid snippets. Taken together, the results argue for CI/CD observability that combines diagnosis, context, and human review rather than treating workflow mining as a stage-classification problem alone.
\end{abstract}

\section{INTRODUCTION}

CI/CD configuration files do more than automate builds. They decide which checks gate a merge, how failures are bounded, when artifacts are published, and what paths lead to deployment. In other words, they encode operational policy. For software-intensive systems, that policy has direct consequences for reliability, security, release cadence, and infrastructure cost. Recent work has shown that LLMs can recover common CI/CD stages directly from raw workflow files at scale \cite{chomatek2025}. Useful as that result is, it still leaves the main engineering questions open: Which workflows look fragile? What counts as normal for a given ecosystem? Which fixes are worth a maintainer's time?

We study those questions by treating CI/CD workflows as analyzable infrastructure rather than passive configuration text. Our pipeline begins with a large GitHub corpus, enriches repositories with language and domain metadata, scans workflows for anti-patterns and stage structure, and then generates repository-level recommendations with several prompting strategies. The goal is not to replace maintainers with an autonomous fixer. The goal is to give them a clearer picture of operational risk and a better starting point for intervention.

The dataset is large enough to make those comparisons meaningful. We scraped 59,550 repositories with at least 1,000 GitHub stars, found CI/CD configurations in 34,225 repositories, and downloaded 127,559 workflow files spanning GitHub Actions, Travis CI, CircleCI, Jenkins, GitLab CI/CD, AppVeyor, and Azure Pipelines. This study is organized around three research questions:
\begin{enumerate}
    \item (RQ1) Which CI/CD anti-patterns dominate large-scale open-source workflows?
    \item (RQ2) How do stage usage and optimization features vary across languages and project domains?
    \item (RQ3) Which prompting strategy yields the most usable automated CI/CD recommendations?
\end{enumerate}

\begin{figure}[t]
    \centering
    \resizebox{\columnwidth}{!}{%
    \begin{tikzpicture}[
        node distance=1.05cm and 0.6cm,
        box/.style={draw, rounded corners, align=center, minimum width=2.65cm, minimum height=1.0cm, fill=blue!6},
        result/.style={draw, rounded corners, align=center, minimum width=2.65cm, minimum height=1.0cm, fill=green!8},
        arrow/.style={-{Latex[length=2.2mm]}, thick}
    ]
        \node[box] (data) {GitHub Corpus\\59,550 repos\\127,559 configs};
        \node[box, right=of data] (enrich) {Metadata Enrichment\\topics + domain labels\\32,513 repos};
        \node[box, right=of enrich] (rq1) {RQ1 Detector\\75,201 configs\\anti-pattern JSON};
        \node[box, below=of rq1] (rq2) {RQ2 Detector\\59,906 configs\\stages + triggers};
        \node[box, left=of rq2] (rq3) {RQ3 Generator\\34k repos/strategy\\YAML repairs};

        \node[result, above=0.9cm of rq1] (out1) {Issue distributions\\top categories\\domain joins};
        \node[result, below=0.9cm of rq2] (out2) {Language significance\\domain stage profiles};
        \node[result, below=0.9cm of rq3] (out3) {Prompting comparison\\validity analysis};

        \draw[arrow] (data) -- (enrich);
        \draw[arrow] (enrich) -- (rq1);
        \draw[arrow] (enrich) -- (rq2);
        \draw[arrow] (enrich) -- (rq3);
        \draw[arrow] (rq1) -- (out1);
        \draw[arrow] (rq2) -- (out2);
        \draw[arrow] (rq3) -- (out3);
        \draw[arrow] (rq1.south west) -- ++(-0.35,-0.35) |- (rq3.north east);

        \node[draw=black!20, rounded corners, fit=(rq1) (rq2) (rq3), inner sep=8pt, label={[align=center]above:LLM analysis pipeline}] {};
    \end{tikzpicture}%
    }
    \vspace{1mm}
    \caption{Study overview. The artifact extends CI/CD analysis from coarse stage recognition to a connected pipeline that enriches repositories, mines anti-patterns and stages, and generates actionable repair suggestions.}
    \label{fig:overview}
\end{figure}
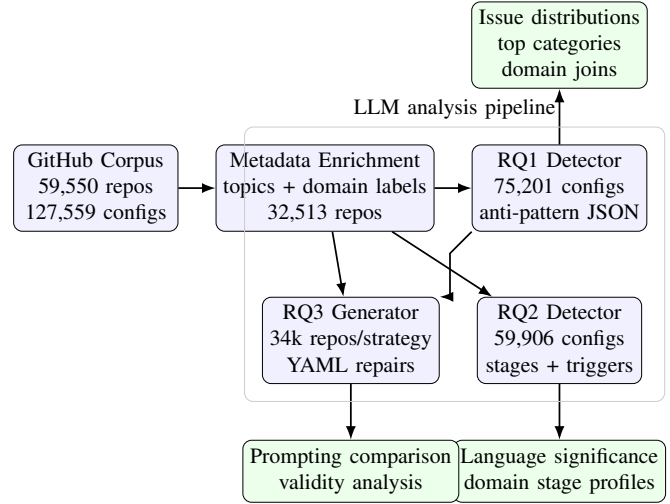

Figure~\ref{fig:overview} summarizes the full workflow, from repository collection and enrichment through anti-pattern detection, stage mining, and recommendation generation. Three findings stand out. First, routine operational debt is everywhere: across 75,201 workflow files, the anti-pattern detector reports 434,769 findings, with reliability issues alone accounting for 150,230 cases. Second, while build and test remain the dominant stages across 59,906 workflows, stage usage changes measurably across language ecosystems and repository domains. Third, for repository-level recommendations, few-shot prompting gives the best overall balance of coverage and machine-checkable validity, whereas iterative prompting shifts more of the output toward critical issues. Taken together, these results suggest that CI/CD governance benefits more from context-aware analysis than from a single universal checklist.

This paper makes three contributions:
\begin{enumerate}
    \item An end-to-end LLM pipeline for CI/CD analysis that combines metadata enrichment, anti-pattern detection, stage mining, and automated recommendation generation in one reusable workflow.
    \item Large-scale empirical evidence from more than 75k analyzed workflow files showing that common CI/CD weaknesses and stage patterns vary systematically with repository context.
    \item A repository-scale comparison of four prompting strategies for machine-assisted CI/CD improvement, making the tradeoff between broad recommendation coverage and severe-issue prioritization explicit.
\end{enumerate}

\section{RELATED WORK}

\subsection{Mining CI/CD Practice at Scale}

Empirical work on CI/CD has moved steadily from anecdotal practice reports to ecosystem-scale measurement. Early studies linked continuous integration adoption on GitHub to collaboration patterns and software-quality outcomes \cite{vasilescu2015,yu2016,hilton2016usage,soares2023}. Later work showed that CI changes surrounding engineering activity and that implementations differ across organizational settings \cite{zhao2017impact,stahl2014modeling}. Subsequent studies examined CI monitoring \cite{santos2024}, CI/CD in cyber-physical systems \cite{zampetti2023}, developer discussions of CI pain points \cite{ouni2023}, and migration between CI services on GitHub \cite{golzadeh2022}. Broader reviews synthesized adoption practices, tools, and recurring challenges in continuous integration, delivery, and deployment \cite{shahin2017continuous,rodriguez2017cdmapping}. More recent GitHub Actions work has focused on workflow structure, adoption effects, outdated dependencies, and maintenance burden \cite{chen2021actions,wessel2023actionsimpact,decan2023outdated,valenzuela2024maintenance}. The closest precursor to our paper is the work of Chomatek \emph{et al.}, who showed that LLMs can infer CI/CD stages directly from configuration files \cite{chomatek2025}. We start from that capability, but ask a broader question: what can we learn once stage recognition is only one component of a larger analysis loop?

\subsection{Security, Reliability, and Misconfiguration Analysis}

CI/CD workflows have also been studied through narrower diagnostic lenses, especially security, testing, and efficiency. Koishybayev \emph{et al.} characterized security weaknesses in GitHub CI workflows \cite{koishybayev2022}; Benedetti \emph{et al.} proposed automated security assessment for GitHub Actions workflows \cite{benedetti2022actions}; Muralee \emph{et al.} introduced a taint-analysis framework for injection vulnerabilities in GitHub workflows and actions \cite{muralee2023argus}; and Feio \emph{et al.} examined DevSecOps practice with an emphasis on continuous security testing \cite{feio2024devsecops}. Related work cataloged security smells in infrastructure-as-code scripts \cite{rahman2019}, traced CI build failures to testing activity in Travis-based development \cite{beller2017oops}, assembled a broad list of bad CI practices \cite{zampetti2020badci}, and documented practitioner tradeoffs among assurance, security, and flexibility in CI tooling \cite{hilton2017tradeoffs}. Weeraddana \emph{et al.} further showed that dependency-related waste remains a measurable CI efficiency problem \cite{weeraddana2024waste}. Our paper is motivated by the same need for structured diagnosis, but we widen the scope: instead of isolating one failure mode, we analyze security, reliability, performance, and maintainability together so that ordinary operational debt is visible alongside explicitly security-focused issues.

\subsection{LLMs for Software and Infrastructure Analysis}

The broader LLM-for-software-engineering literature shows that foundation models can already reason over workflow and testing artifacts with useful accuracy, and recent surveys cover that trend across software engineering tasks \cite{zhangsurvey2023}. Zhang \emph{et al.} examined LLM effectiveness on GitHub workflows directly \cite{zhang2024}; Li \emph{et al.} and Feldt \emph{et al.} studied testing-oriented applications and autonomous testing agents \cite{li2025,feldt2023}. Our work differs less in model choice than in evaluation posture: we connect enrichment, diagnosis, and recommendation generation in a single infrastructure-facing pipeline, and evaluate machine-checkable output quality at repository scale rather than stopping at task accuracy on a small benchmark.

\subsection{Variation Across Software Project Types}

Software process norms vary by ecosystem. Studies of testing adoption \cite{kochhar2013}, peer review \cite{rigby2013}, and release engineering \cite{adams2016} have all found that repository context shapes what developers treat as standard practice. We follow that thread into CI/CD workflows, measuring how stage presence and optimization features differ across language communities and repository types.

\section{DATASET AND PIPELINE}

\subsection{Dataset Construction}

We built a local scraping and analysis framework. The scraper queries GitHub for repositories with at least 1,000 stars, inspects their trees for CI/CD files, and downloads supported configurations. That produced 59,550 repositories in total---34,225 with CI/CD and 127,559 downloaded configuration files.

Repository metadata was enriched in two stages: topic tags pulled from the GitHub API, and a domain classifier that assigned each repository to one of seven categories---\texttt{library}, \texttt{web\_app}, \texttt{ml\_data\_science}, \texttt{mobile\_app}, \texttt{devops\_infra}, \texttt{cli\_utility}, or \texttt{other}. The enrichment pass covers 32,513 repositories, with Python, TypeScript, JavaScript, Go, and C++ as the largest language groups. Table~\ref{tab:dataset} summarizes dataset size and coverage per research question.

\begin{table}[t]
\centering
\caption{Dataset and analysis coverage. RQ1 and RQ2 operate at configuration-file level; RQ3 operates at repository level.}
\label{tab:dataset}
\begin{tabular}{lr}
\toprule
\textbf{Artifact} & \textbf{Count} \\
\midrule
Scraped repositories ($\geq$1,000 stars) & 59,550 \\
Repositories with CI/CD detected & 34,225 \\
Downloaded CI/CD configuration files & 127,559 \\
Domain-enriched repositories & 32,513 \\
RQ1 anti-pattern analyses & 75,201 \\
RQ2 stage analyses & 59,906 \\
RQ3 zero-shot repositories & 34,152 \\
RQ3 few-shot repositories & 33,916 \\
RQ3 retrieval-augmented repositories & 34,146 \\
RQ3 iterative repositories & 34,123 \\
\bottomrule
\end{tabular}
\end{table}

\subsection{Research Posture}

The pipeline does two things: measures software delivery infrastructure and suggests improvements to it. Repository metadata provides context, workflow analysis captures the current operational state, and recommendation generation proposes interventions. These stages produce different kinds of evidence, so we evaluate them separately. Corpus counts describe what exists in the data. Structured model outputs summarize what the detector reports. Parser-based checks test whether generated snippets are at least machine-formed. That separation matters because it keeps us precise about what the paper actually claims --- and what it does not.

\subsection{RQ1: Anti-Pattern Detection}

For RQ1, we run an LLM-based anti-pattern detector on each CI/CD configuration file. The prompt instructs the model to act as a senior DevOps auditor and return a structured list of anti-patterns, each tagged with dimension, category, severity, remediation, and confidence. Findings fall under four dimensions: security, reliability, performance, and maintainability. Outputs are stored as JSONL and summarized into counts by dimension and severity.

The goal is comparative measurement, not exhaustive manual adjudication. The counts should be read as diagnostic signals from a consistent analysis procedure---they reflect what the detector finds, not how often these anti-patterns actually appear across open-source CI/CD in general.

\subsection{RQ2: Stage Detection and Practice Analysis}

For RQ2, we apply an LLM-based workflow-structure analyzer to each configuration file. It emits stage labels, tool mentions, trigger events, job counts, and binary indicators for matrix builds and caching. A statistics script aggregates those outputs and runs a $\chi^2$ test on the language-by-stage contingency table for the ten most frequent languages, with Cramer's $V$ as the effect-size estimate.

We also join stage outputs back to the domain-enrichment results by repository. That lets us compare workflows against their repository context rather than collapsing everything into one global frequency table.

\subsection{RQ3: Recommendation Generation}

For RQ3, we move to the repository level so multiple workflows can be considered together. Each recommendation has a type (\texttt{fix}, \texttt{add}, or \texttt{optimize}), a priority level, an explanation, and a YAML snippet. We compare four prompting strategies: zero-shot, few-shot with exemplars, retrieval-augmented, and iterative with a critique-and-revise loop.

The evaluation is fully automated. For each strategy, we report recommendation counts, type and priority distributions, and YAML validity via a parser-based check---a consistent basis for comparison when expert review is not available for every output.

\section{RESULTS}

\subsection{RQ1: Anti-Patterns Are Frequent and Reliability-Dominated}

RQ1 analyzed 75,201 CI/CD configurations and found 434,769 anti-patterns, averaging $434769/75201 \approx 5.78$ findings per configuration. Reliability is the biggest issue family with 150,230 findings, followed by maintainability (114,412), performance (91,091), and security (79,036). Most findings are in the center of the severity scale, yet the detector reports 94,214 high-or-critical cases. The bottom line is clear: operational debt in CI/CD is prevalent even in highly visible open-source repositories.

\begin{figure}[t]
    \centering
    \begin{minipage}[t]{0.48\textwidth}
        \centering
        \begin{tikzpicture}
            \begin{axis}[
                ybar,
                ymin=0,
                ylabel={Issues (thousands)},
                symbolic x coords={Reliability,Maintain.,Performance,Security},
                xtick=data,
                x tick label style={rotate=20, anchor=east, font=\small},
                width=\textwidth,
                height=5.0cm,
                bar width=13pt,
                nodes near coords,
                nodes near coords align={vertical},
                every node near coord/.append style={font=\scriptsize},
                enlarge x limits=0.15
            ]
            \addplot[fill=blue!45] coordinates {(Reliability,150.2) (Maintain.,114.4) (Performance,91.1) (Security,79.0)};
            \end{axis}
        \end{tikzpicture}
        
        \vspace{1mm}
        {\footnotesize\emph{(a) Issue counts by anti-pattern dimension.}}
    \end{minipage}
    \hfill
    \begin{minipage}[t]{0.48\textwidth}
        \centering
        \begin{tikzpicture}
            \begin{axis}[
                ybar,
                ymin=0,
                ylabel={Issues (thousands)},
                symbolic x coords={Critical,High,Medium,Low,Info},
                xtick=data,
                x tick label style={rotate=20, anchor=east, font=\small},
                width=\textwidth,
                height=5.0cm,
                bar width=11pt,
                nodes near coords,
                nodes near coords align={vertical},
                every node near coord/.append style={font=\scriptsize},
                enlarge x limits=0.12
            ]
            \addplot[fill=orange!55] coordinates {(Critical,14.8) (High,79.4) (Medium,188.4) (Low,144.5) (Info,7.7)};
            \end{axis}
        \end{tikzpicture}
        
        \vspace{1mm}
        {\footnotesize\emph{(b) Issue counts by severity.}}
    \end{minipage}
    \caption{Global RQ1 anti-pattern distribution across 75,201 analyzed configuration files and 434,769 total findings.}
    \label{fig:rq1-distribution}
\end{figure}
The breakdown by category illustrates the specifics of that debt. Figure~\ref{fig:rq1-distribution} represents the distributions by dimension and by severity. After consolidation of near-duplicate detector labels for the same issue family, the most frequently reported issues are missing timeout configurations (58,414 findings), followed by missing caching (29,987), unpinned third-party actions (27,925), and duplicated workflow configuration (17,692). These are not obscure edge cases. They demonstrate that numerous projects still leave failure boundaries loose, repeat work that could be avoided, and similarly accrued configuration as workflows expand over time.
The issue mix is relatively homogeneous across domains, although the findings density is different. Mobile and ML/data-science projects have the highest average counts per configuration (6.10 and 6.08, respectively), while libraries are marginally lower at 5.63. Reliability remains the leading dimension in all domains which reiterates that the prevailing pain points are operationally common before they become exotic.

\subsection{RQ2: CI/CD Practice Varies by Language and Domain}
RQ2 processed 59,906 configuration files. The most frequent stage is build (31,649 configurations), followed by test (22,750), release (14,331), lint (11,616), and containerization (9,519). Only 5,092 configurations have a security scan stage explicitly stated.  The prevalence of optimization features is lower than one might expect; only 28.8\% of configurations use matrix builds, and 24.9\% use caching.

\begin{figure}[t]
    \centering
    \begin{minipage}[t]{0.48\textwidth}
        \centering
        \begin{tikzpicture}
            \begin{axis}[
                ybar,
                ymin=0,
                ylabel={Configurations (thousands)},
                symbolic x coords={Build,Test,Other,Release,Lint,Containerization},
                xtick=data,
                x tick label style={rotate=22, anchor=east, font=\small},
                width=\textwidth,
                height=5.0cm,
                bar width=11pt,
                nodes near coords,
                every node near coord/.append style={font=\scriptsize},
                enlarge x limits=0.12
            ]
            \addplot[fill=teal!55] coordinates {(Build,31.6) (Test,22.8) (Other,21.2) (Release,14.3) (Lint,11.6) (Containerization,9.5)};
            \end{axis}
        \end{tikzpicture}
        
        \vspace{1mm}
        {\footnotesize\emph{(a) Most common detected stages.}}
    \end{minipage}
    \hfill
    \begin{minipage}[t]{0.48\textwidth}
        \centering
        \begin{tikzpicture}
            \begin{axis}[
                ybar,
                ymin=0,
                ymax=40,
                ylabel={Configurations (\%)},
                symbolic x coords={Library,Web app,ML/data,Mobile,DevOps},
                xtick=data,
                x tick label style={rotate=25, anchor=east, font=\small},
                width=\textwidth,
                height=5.0cm,
                bar width=8pt,
                legend style={at={(0.5,1.02)}, anchor=south, legend columns=2, draw=none},
                enlarge x limits=0.16
            ]
            \addplot[fill=purple!45] coordinates {(Library,33.1) (Web app,22.0) (ML/data,30.4) (Mobile,12.4) (DevOps,25.0)};
            \addplot[fill=green!45] coordinates {(Library,25.8) (Web app,29.0) (ML/data,21.6) (Mobile,35.5) (DevOps,20.6)};
            \legend{Matrix,Cache}
            \end{axis}
        \end{tikzpicture}
        
        \vspace{1mm}
        {\footnotesize\emph{(b) Matrix and cache adoption by domain.}}
    \end{minipage}
    \caption{RQ2 summary across 59,906 analyzed configuration files.}
    \label{fig:rq2-summary}
\end{figure}
There is a statistically meaningful difference with stage distribution, at the language level, with a p-value of less than 0.001 ($\chi^2 = 4168.88$). The effect size is modest (Cramer's $V = 0.063$). The general stage frequencies and optimization-feature utilization are Figure~\ref{fig:rq2-summary}. The main takeaway is not that ecosystems need individual taxonomies, but that there are no singular workflow templates that suit every ecosystem. The same overarching stages are present, but their distribution shifts according to the expectations of the ecosystem. Relatively strong testing is seen in Python and Go repositories, Rust has a higher than average adoption of lint, while C\# projects have more than average release workflow activities.

\begin{table}[t]
\centering
\caption{Domain-specific stage prevalence in RQ2. Values are percentages of configurations in each domain.}
\label{tab:rq2-domain}
\begingroup
\footnotesize
\setlength{\tabcolsep}{3pt}
\begin{tabular}{lrrrrrrr}
\toprule
\textbf{Domain} & \textbf{Build} & \textbf{Test} & \textbf{Release} & \textbf{Container} & \textbf{Lint} & \textbf{Sec.} & \textbf{Cache} \\
\midrule
library & 55.6 & 45.7 & 21.3 & 9.4 & 20.5 & 7.9 & 25.8 \\
web\_app & 46.3 & 27.7 & 24.2 & 23.4 & 18.2 & 8.5 & 29.0 \\
ml\_data\_science & 49.6 & 41.5 & 21.6 & 20.4 & 20.1 & 4.9 & 21.6 \\
mobile\_app & 62.6 & 21.6 & 32.0 & 12.8 & 17.1 & 5.7 & 35.5 \\
devops\_infra & 44.6 & 35.7 & 24.1 & 24.8 & 17.6 & 12.7 & 20.6 \\
\bottomrule
\end{tabular}
\endgroup
\end{table}

The domain-level join makes those differences easier to read. Table~\ref{tab:rq2-domain} reports the domain-specific stage breakdown. 
Mobile projects exhibit the most releases (32.0\%) and the most usage of caches (35.5\%), but they have a low test-stage prevalence (21.6\%). Libraries show the greatest adoption of tests (45.7\%) and a decently high frequency of code-quality stages, while DevOps/infrastructure repos are notable for pipelines with a lot of containerization (24.8\%). These contrasts highlight the importance of contextual baselines. In a given domain, a workflow that looks bare might be the opposite in another.

\subsection{RQ3: Few-Shot Prompting Is the Strongest Default}
In our evaluation, few-shot prompting gives the strongest overall recommendation profile. Table~\ref{tab:rq3-strategies} presents the cross-strategy summary. Few-shot achieved the highest overall recommendations (279,811 across 33,916 repos), the highest average recommendations per repo (8.25), and the best YAML validity rate (96.1\%). In terms of output quantity and validity, zero-shot and retrieval-augmented prompting are virtually identical, which suggests that the current retrieval approach does little beyond a solid direct prompt.

\begin{table}[t]
\centering
\caption{Automated comparison of recommendation prompting strategies. Shares are percentages of all recommendations emitted by a strategy.}
\label{tab:rq3-strategies}
\begingroup
\footnotesize
\setlength{\tabcolsep}{3pt}
\begin{tabular}{lrrrrr}
\toprule
\textbf{Strategy} & \textbf{Repos} & \textbf{Avg./repo} & \textbf{Fix share} & \textbf{Critical share} & \textbf{YAML valid} \\
\midrule
Zero-shot & 34,152 & 8.05 & 38.3 & 13.2 & 94.8 \\
Few-shot & 33,916 & 8.25 & 38.3 & 10.6 & 96.1 \\
RAG & 34,146 & 8.06 & 38.5 & 13.2 & 94.8 \\
Iterative & 34,123 & 7.24 & 40.3 & 16.7 & 95.5 \\
\bottomrule
\end{tabular}
\endgroup
\end{table}

Iterative prompting demonstrates a different pattern of behavior. Overall, fewer suggestions are made (7.24 suggestions on average per repository), however, 41,276 are noted as critical. It suggests that the critique-and-revise loop acts more like a prioritization filter than a breadth amplifier. If the goal is wide coverage with syntactically valid repair suggestions, few-shot is the better default. If the goal is narrower triage around severe issues, iterative prompting becomes more appealing.

The domain pattern for RQ3 largely aligns with RQ2. Using few-shot prompting, the average number of suggestions for DevOps/infrastructure projects is the highest (9.03 suggestions per repository), followed by web (8.64) and mobile (8.59) applications. Among the major domains, libraries receive the least (8.01), which corroborates the previous claim that their workflows are comparatively more regular and test-heavy.

\section{DISCUSSION}

The most striking result is that the dominant problems are ordinary ones. Missing timeouts, missing caching, duplicated configuration, and unpinned actions account for a large share of the detector output. These are not obscure failure modes, but basic operational controls. The result suggests that for a workflow-intelligence system that is designed to help real maintainers, routine hygiene deserves at least as much attention as rare security anomalies.

The stage analysis also makes a clear second point: CI/CD practice is contextual. The language and domain-level differences are not so large that every ecosystem needs its own theory of CI/CD, but they are large enough to undermine a single global checklist. In practice, recommendation quality should improve when workflows are judged relative to comparable projects rather than against corpus-wide averages alone.

From the cyber-systems engineering perspective, the pipeline forms a simple sensing-analysis-decision loop over software delivery infrastructure. Workflow files and repository metadata provide the observed state; the mining modules turn that state into structured diagnostics; and the recommendation module proposes changes that a maintainer can inspect. The purpose of the loop is not full autonomy. It is to make human review better informed and more focused.

Above interpretation is reinforced by the prompting comparison. Prompt choice changes not just how much text the model generates, but what kind of assistance it provides. Few-shot prompting broadens coverage and preserves syntactic validity, while iterative prompting concentrates more heavily on severe fixes. Making that tradeoff explicit is more useful than presenting one strategy as universally best.

The results have several limitations. Firstly, the corpus is GitHub-centric and biased toward highly starred repositories. Domain enrichment covers 32,513 repositories rather than the entire CI/CD-positive set. RQ1 and RQ3 also rely on scalable automated evaluation rather than expert-judged usefulness, and the reported outputs come from one configured model family. The comparative conclusions hold in the paper despite those constraints: common issue families are stable enough to measure, workflow structure varies with context, and prompting strategy materially changes the kind of recommendations the system produces.

\section{CONCLUSION}
We presented an LLM-based pipeline for analyzing CI/CD workflows over a large GitHub corpus. Instead of stopping at stage detection, the anti-pattern reporting, contextual workflow mining, workflow level recommendations, and repository level recommendations are integrated. The empirical results show that the detector output is dominated by issues pertaining to reliability and maintainability, that workflow structure shifts systematically with repository language and domain, and that default few-shot prompting is the strongest when the target is broad, machine-checkable, and recommendable quality.

The most effective next step is the external validation concerning RQ1, proposed targeted human annotation, RQ3 proposed downstream studies concerning real pull request workflows, and proposed expert evaluation of the recommendation's utility. These additions would firm up the practical justification for applying this sort of CI/CD analysis to human-centered governance practices.

\section*{APPENDIX}

Table~\ref{tab:appendix-language-stage} reports language-specific stage prevalence for the ten largest language groups in the RQ2 outputs. Table~\ref{tab:appendix-top-categories} lists the most common anti-pattern categories across RQ1 after merging near-duplicate detector labels.

\begin{table}[t]
\centering
\caption{Stage prevalence for the ten largest language groups in RQ2.}
\label{tab:appendix-language-stage}
\begin{tabular}{lrrrrr}
\toprule
\textbf{Language} & \textbf{Configs} & \textbf{Build} & \textbf{Test} & \textbf{Lint} & \textbf{Release} \\
\midrule
Python & 10,208 & 47.1 & 42.0 & 20.7 & 24.1 \\
TypeScript & 9,357 & 54.1 & 32.0 & 21.1 & 27.8 \\
Go & 7,330 & 42.9 & 38.0 & 20.8 & 24.3 \\
C++ & 4,684 & 68.0 & 39.0 & 13.0 & 23.6 \\
JavaScript & 4,264 & 48.9 & 43.7 & 20.6 & 19.9 \\
Rust & 3,854 & 56.9 & 42.6 & 29.0 & 28.2 \\
Java & 3,486 & 62.2 & 42.6 & 10.6 & 21.3 \\
C & 2,527 & 71.6 & 43.1 & 12.4 & 20.3 \\
C\# & 2,011 & 64.9 & 30.9 & 7.1 & 34.8 \\
PHP & 1,710 & 38.3 & 41.8 & 32.2 & 14.6 \\
\bottomrule
\end{tabular}
\end{table}

\begin{table}[t]
\centering
\caption{Most common anti-pattern categories in RQ1 after merging near-duplicate detector labels for the same issue family.}
\label{tab:appendix-top-categories}
\begin{tabular}{lll}
\toprule
\textbf{Dimension} & \textbf{Category} & \textbf{Count} \\
\midrule
reliability & \texttt{timeout\_configuration} & 58,414 \\
performance & \texttt{missing\_caching} & 29,987 \\
security & \texttt{unpinned\_action} & 27,925 \\
maintainability & \texttt{duplicated\_configuration} & 17,692 \\
maintainability & \texttt{hardcoded\_values} & 15,052 \\
reliability & \texttt{error\_handling} & 14,270 \\
performance & \texttt{unnecessary\_full\_checkout} & 8,229 \\
reliability & \texttt{missing\_concurrency\_control} & 4,939 \\
\bottomrule
\end{tabular}
\end{table}

\bibliographystyle{IEEEtran}
\bibliography{references}

\end{document}